\documentclass[a4paper,11pt]{article}
\usepackage{amsmath,amsthm,amssymb,float,graphicx}
\newcommand{\nc}{\newcommand}
\nc{\rnc}{\renewcommand}
\nc{\nn}{\nonumber}
\nc{\der}{{\partial}}
\rnc{\Im}{{\textrm{Im}\,}}
\rnc{\Re}{{\textrm{Re}\,}}
\nc{\db}{\displaybreak[0]\\}
\nc{\bra}{\langle}
\nc{\ket}{\rangle}
\nc{\astl}{\underset{\mathcal{L}_l}{\ast}}
\nc{\astj}{\underset{\mathcal{L}_j}{\ast}}
\nc{\astll}{\underset{\mathcal{L}_1}{\ast}}
\nc{\Yt}{Y^{\textrm{th}}}
\nc{\rmd}{{\rm d}}
\nc{\rmi}{{\rm i}}
\nc{\rme}{{\rm e}}

\nc{\End}{\mathrm{End}}

\begin{document}

\title{Can a CNN trained on the Ising model detect the 
phase transition  of the $q$-state Potts model?}

\author{
Kimihiko Fukushima\thanks{E-mail: 2213091@alumni.tus.ac.jp} \, and
Kazumitsu Sakai\thanks{E-mail: k.sakai@rs.tus.ac.jp} \,
\\\\
\textit{Department of Physics,
Tokyo University of Science,}\\
 \textit{Kagurazaka 1-3, Shinjuku-ku, Tokyo, 162-8601, Japan} \\
\\\\
\\
}

\date{April 8, 2021}
\maketitle

\begin{abstract}
Employing a deep convolutional neural network (deep CNN) trained on spin configurations 
of the 2D Ising model and the temperatures, we examine whether the deep
CNN can detect the phase transition of the 2D $q$-state Potts model.
To this end, we generate binarized images of spin configurations
of the $q$-state Potts model ($q\ge 3$) by replacing the spin variables 
$\{0,1,\dots,\lfloor q/2\rfloor-1\}$ and $\{\lfloor q/2\rfloor,\dots,q-1\}$
with  $\{0\}$ and $\{1\}$, respectively. Then, we
input these images to the trained CNN to output the predicted 
temperatures. The binarized images of the $q$-state Potts model 
are entirely different from Ising spin configurations, particularly at the 
transition temperature. Moreover, our CNN model
is not trained on the information about whether
phases are ordered/disordered   but is naively trained by Ising spin 
configurations labeled with temperatures at which they are generated.
Nevertheless, the deep CNN can detect the transition 
point with high accuracy, regardless of the type of transition. 
We  also find that, in the high-temperature region, the CNN outputs the temperature 
based on the internal energy, whereas, in the low-temperature region, the output
depends on the magnetization and possibly the internal energy as well. 
However, in the vicinity of the transition point, the CNN may use more general 
factors to detect the transition point.
\end{abstract}
\vspace*{1cm}
\noindent
1. {\it \large Introduction}\,
Machine learning (ML) employing an artificial neural network (NN)
has seen renewed interest in recent years  and has been widely applied in various branches
of science using its ability to capture features and classify them
 (see, e.g.,  Refs. \cite{GBC16,LBH}).
A NN is conceptually inspired by the structure of the brain neurons: it consists 
of a network of nodes (representing neurons) aligned in layers so 
that the nodes in each layer are connected by the links (synapses)
through which  data are passed to each node.
The node is activated (fired) when the sum of the weighted data exceeds 
a threshold value called bias. 
These weights (corresponding to the strengths of synaptic connections)
are at first randomly initialized, and trained  by repeatedly passing the training data 
through the network until the NN outputs something meaningful.
A NN trained in this way can have the ability to solve complex 
real-world problems that conventional approaches have failed to handle, 
such as object recognition and detection in images. There exist different types of
NNs for different tasks. Among them, a convolutional neural network (CNN), which 
will be used in this study, is particularly suitable for processing 2D data such as images
(see Ref. \cite{RW} for a recent review).

It is natural that NNs, which are capable of extracting  specific features 
of real-world objects and classifying them, have begun to be used as  tools 
in studies of theoretical physics (see, e.g., Refs. \cite{CCCDSTV,TTH}).
Let us focus our attention on an application of NNs for problems of order--disorder 
phase transitions \cite{CM17,CMK17,LTJ18,KK18,BAL20,CAAL20,YS20,SMOL20,
TT17,AOT18,KKT19,MT18,LW19,
Wang16,NLH17,Wetzel17,MM18,CFMPRC19,AACP20,WTJ20,DB20}, which is  the main topic 
considered in this letter. More specifically, NNs with supervised \cite{CM17,CMK17,LTJ18,KK18,BAL20,CAAL20,YS20,SMOL20,TT17,AOT18,KKT19,MT18,LW19} and unsupervised
 \cite{Wang16,NLH17,Wetzel17,MM18,CFMPRC19,AACP20,WTJ20,DB20} learning
have been used to accurately identify phases and phase transitions. 
In supervised learning, each training data (e.g., an image of a spin configuration) 
is supplemented by some labels (e.g., ordered/disordered, temperature,  magnetization, etc.), 
and the NN is trained until it outputs values approximately identical to the 
labels of the input data. In unsupervised learning, on the other hand, 
the training data are supplemented by no such labels, and the NN is trained until 
it extracts some discriminative features from the input data. 

In supervised learning, there exist several approaches to detect transition temperatures.
(i) The first approach relies on a binary classification \cite{CM17,CMK17,LTJ18,KK18,
BAL20,CAAL20,YS20}: a NN is trained on a dataset where each data is labeled with, e.g., 
$1$ (ordered) or $0$ (disordered) so that it identifies whether the input is ordered 
or disordered.  Namely, prior information about phases is 
provided to the NN as labels of training samples. The trained NN can precisely 
detect the transition temperature by automatically reading the value of the 
order parameter of the input sample.
(ii) The second approach is more naive and trains a NN only on  data 
labeled with the temperatures at which they are generated 
\cite{TT17,AOT18,KKT19,MT18}.  In other words, information about phases is not
provided to the NN. Nevertheless, the NN spontaneously captures phase transitions
by automatically detecting the internal energy or magnetization \cite{KKT19}.
(iii) The third approach utilizes a transferability of NNs. 
A NN trained on one system surprisingly detects the transition point in 
other similar systems obtained by, e.g., changing a lattice topology or
a form of interaction in the Ising model \cite{CM17,CFMPRC19,CAAL20}, 
a filling number in 
the Hubbard model \cite{CMK17}, and 
a spin state number in the $q$-state Potts model \cite{SMOL20,BAL20,YS20}.

Despite these successes, in many cases, NNs work as black boxes, and a 
systematic understanding of how NNs identify and extract features from complex objects 
(in our case, images of spin configurations) is a crucial ingredient for the 
universal application of NNs to various problems in the natural/social sciences.
See Ref. \cite{ISY18} for an interesting argument in terms of some
relation between renormalization groups and deep NNs.

In this letter, combining the second and third approaches described above, 
we examine whether a deep CNN (deep CNN) trained on 
the Ising model can identify phase transitions of the $q$-state Potts models.
Moreover, analyzing relations between latent variables and physical quantities,
we consider how the CNN detects the phase transition of the $q$-state Potts models.
As mentioned above, the transferability of an Ising-trained NN to the
Potts models has also been  investigated in Refs. \cite{SMOL20,BAL20,YS20} by a binary 
classification or, equivalently, by providing  prior information about phases.  
We would like to emphasize that our CNN model is not trained on the information about
whether phases are ordered/disordered  but is naively trained by Ising spin configurations 
labeled with the
temperatures at which they are generated, and our Ising-trained CNN is  easily
applicable for detection of the transition point of the $q$-state Potts model.
Though the physical properties of the Ising model and 
the $q$-state Potts model with $q\ge 3$ are  different 
especially at the transition temperature, the CNN can accurately predict the transition 
temperature, regardless of the type of transition. \\

\begin{figure}[tpbp]
\centering
    \includegraphics[width=0.8\textwidth]{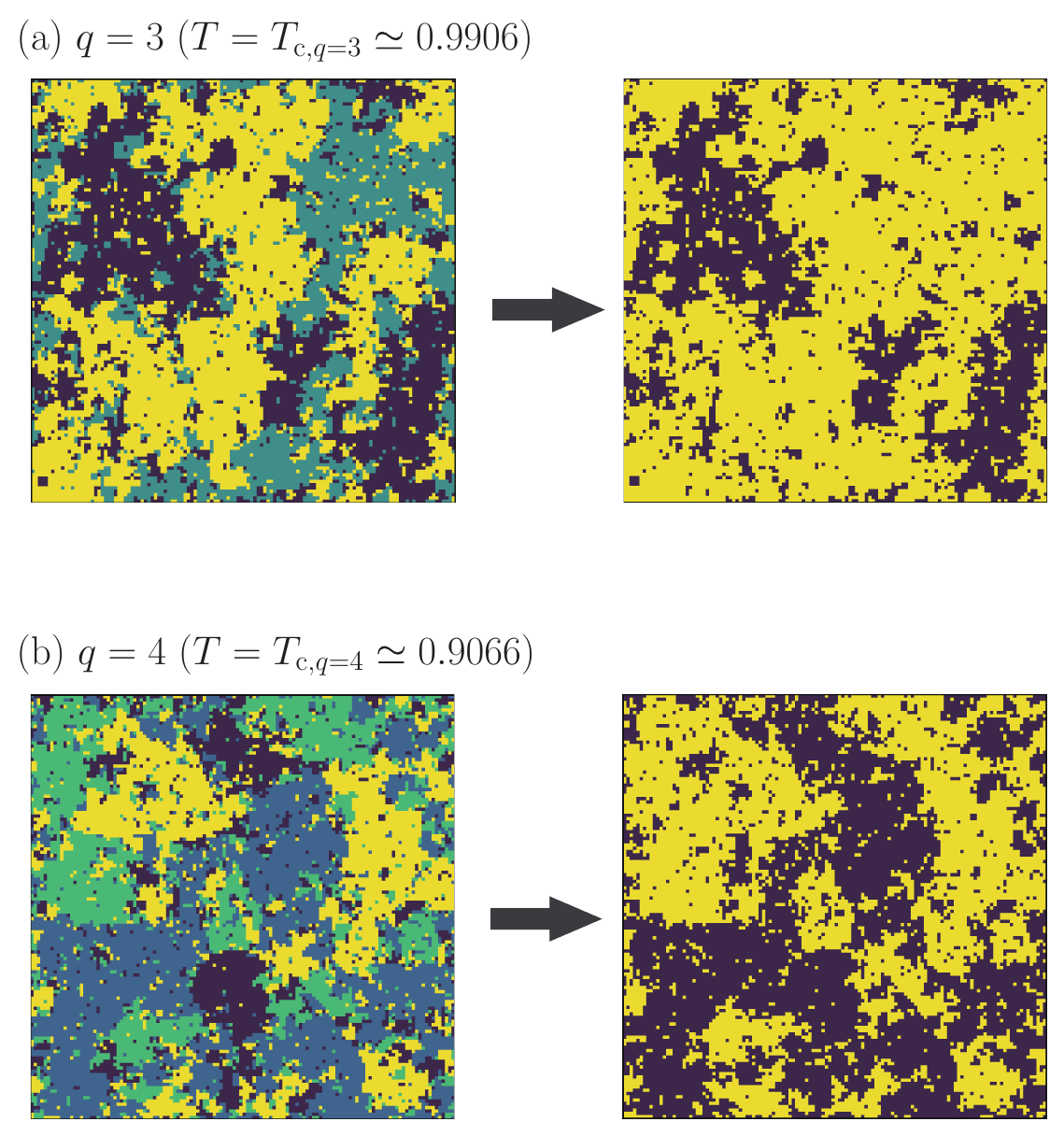}
    \caption{Images of spin configurations for the $3$- and $4$-state
Potts models at criticality (left panel) and their binarized images
(right panel). The critical temperatures $T_{{\rm c}, q}$ are calculated
from the behaviors of the internal energies. 
The spin configurations are generated on a square lattice 
of size $128\times 128$ with free boundary conditions.
(a) Left: An image of a spin configuration of the $3$-state Potts model
at criticality. 
The spin variables $0$, $1$, and $2$ 
are colored purple, green and yellow, respectively. 
Right: The binarized
image obtained by replacing the variables $\{0\}$ and $\{1,2\}$ 
with $\{0\}$ and $\{1\}$, respectively. The spin variables $0$ and $1$
are colored purple and yellow, respectively. 
(b) Left: An image of a spin configuration of the $4$-state Potts model
at criticality. The spin variables $0$, $1$, $2$ and 
$3$ are colored purple, blue, green and yellow, respectively. 
Right: The binarized
image obtained by replacing the variables $\{0,1\}$ and $\{2,3\}$ 
with $\{0\}$ and $\{1\}$, respectively. }
    \label{binarized}
\end{figure}

\noindent
2. {\it \large $q$-state Potts model}\,
Let us summarize the $q$-state Potts model, concentrating on the 
phase transition. The model on an $L\times L$ square lattice is defined by
\begin{equation}
    H(\{\sigma_j\}) = -J \sum_{\langle j,k \rangle}\delta_{\sigma_j,\sigma_k},
\quad \sigma_j \in \{0,1,\dots, q-1\}, \label{Potts}
\end{equation}
where $\bra j, k \ket$ denotes nearest-neighbor pairs. We shall exclusively consider the 
ferromagnetic model $J>0$ and set $J=k_{\rm B}=1$ ($k_{\rm B}$: the Boltzmann 
constant) for convenience.  For $q = 2$, the model is equivalent
to the Ising model. In the thermodynamic limit $L\to\infty$,
the model undergoes a second-order phase transition for $q\le 4$, while for
$q> 4$ it undergoes a first-order transition \cite{Wu1982} at 
the transition temperature $T=T^{\rm \infty}_{\rm c}(q)$:
\begin{equation}
T^{\infty}_{\rm c}(q)=1/\log(1+\sqrt{q}).
\label{transition}
\end{equation}  
In particular, for $q\le 4$,
 each scaling behavior is classified into a {\it different} 
universality class: the critical properties for 
the $q=2,3,4$ Potts models are characterized by conformal field theory (CFT) 
with the central charge $c=1/2,4/5,1$, respectively \cite{VF}.

\begin{figure}[ttt]
\centering
    \includegraphics[width=0.95\textwidth]{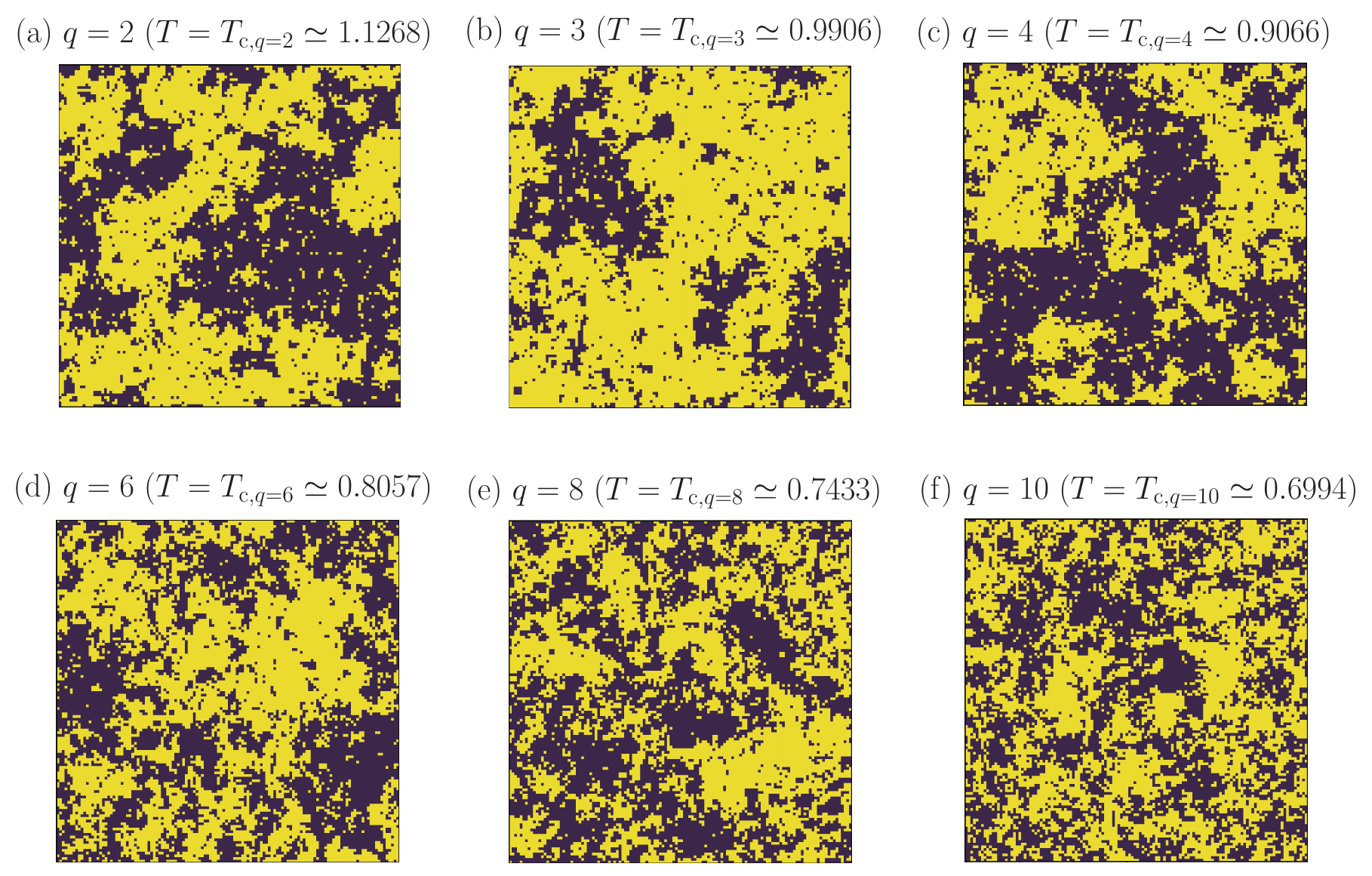}
    \caption{Binarized images of spin configurations at the transition points for 
the various spin states (see also Fig.~\ref{binarized}). 
For comparison, an image for the Ising model ($q=2$) 
is also depicted in (a). The geometric properties of these binarized images are  
different for each. For instance, the fractal dimensions $d_{\rm f}$ of the spin-cluster 
boundaries for $q=2$, $q=3$, and $q=4$
are  given by $d_{\rm f}=11/8$, $d_{\rm f}=17/12$, and $d_{\rm f}= 3/2$, respectively. }
    \label{binarized2}
\end{figure}

Our purpose is to examine whether the Ising-trained CNN can detect the phase transition
of the $q$-state ($q\ge 3$) Potts model. The trained CNN, however, 
may only adapt to classifying the binary images: we must appropriately binarize the 
images of spin configurations of the $q$-state Potts model while loosing as few 
of the essential properties of the phase transition as possible.
The simplest way is to divide the spin variables in the given configuration
into two parts $\{0,1,\dots,\lfloor q/2\rfloor-1\}$ and $\{\lfloor q/2\rfloor,\dots,q-1\}$
and replace them with $\{0\}$ and $\{1\}$, respectively (see Fig.~\ref{binarized} for 
the $q=3$
and $q=4$ cases). Let us denote the resultant 
binarized configuration  by $\{\sigma'_j\}$ ($\sigma'_j\in\{0,1\}$). Also, we define
the internal energy $E$ and the magnetization $M$
for the transformed model as
\begin{equation}
E:=\frac{H(\{\sigma_j'\})}{L(L-1)}, \quad M:=\frac{1}{L^2}\sum_j(2\sigma'_j-1),
\label{internal}
\end{equation}
respectively. We stress that,
under this transformation,
the transition temperature and the  transition type 
(i.e., first/second) are invariant in the thermodynamic limit. 
Furthermore, some geometric properties at criticality are still retained in 
the binarized models. For instance, the fractal dimensions $d_{\rm f}$ of
the cluster boundaries of the binarized models for $q=3$ and $q=4$ are, respectively,
given by $d_{\rm f}=17/12$ \cite{GC} and $d_{\rm f}= 3/2$ \cite{FS}, 
which are consistent with the prediction of CFT. In Fig. 2, we depict binarized images 
at the transition points for various spin states.\\

\begin{figure}[thbp]
\centering
    \includegraphics[width=0.8\textwidth]{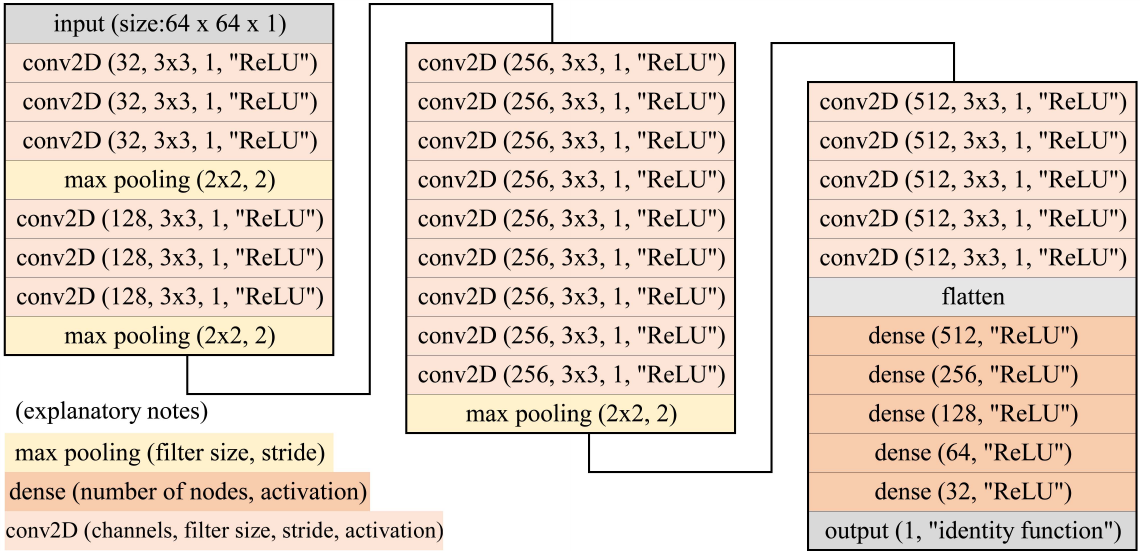}
    \caption{A schematic representation of the deep CNN model used in the study. The CNN consists
of 20 convolutional layers and 5 fully connected layers.  Batch normalization is also 
applied after the input layer and each ReLU activation.}
    \label{cnn}
\end{figure}

\noindent
3. {\it \large Ising trained deep CNN}\,
Our procedure goes as follows. (i) First, we generate a deep CNN by training it on 
a dataset consisting of pairs of images of Ising spin configurations 
and the temperatures. 
More specifically, using the Wolff algorithm \cite{Wolff}, we generate $6,000$ images 
on a $128\times 128$ square lattice with free boundary conditions
for each temperature ranging 
from $T_{{\rm c},q=2}^{\infty}-0.5$ to $T^{\infty}_{{\rm c},q=2}+0.5$ 
($T^{\infty}_{{\rm c},q=2}=1/\log(1+\sqrt{2})\simeq 1.1346$)
in increments of 0.01. To eliminate boundary effects and also reduce the 
processing time, we actually use  images of size $64\times 64$, which are cropped 
from the center of the corresponding original images. 
(ii) Then, as input images, we prepare binarized images of spin configurations of the
$q$-state Potts model in a method similar to that for the training images: 
generate the images of size $128\times128$ using the Wolff algorithm, 
binarize them as explained above, and  then crop the images to
$64\times64$ from the center of the original images.
(iii) Finally, we input the binarized  images to the trained CNN to output 
the predicted temperatures.

Our deep CNN model  comprises  20 convolutional layers and 5 fully connected layers, 
as pictorially depicted in Fig.~\ref{cnn}, and is designed to be somewhat deeper than an
ordinary CNN 
so as to increase the accuracy of the output, the versatility, and the flexibility 
in learning. In each convolutional layer, an image is convolved with $3\times3$ filters, 
a stride of 1, and padding of 0 s.
In each max pooling layer, the filter size and the stride, respectively, are set to $2 \times  2$ 
and $2$. 
To prevent overfitting and vanishing of a gradient, batch normalization \cite{Ioffe2015} is applied after the input layer and each ReLU (Rectified Linear Unit) activation. 
An identity activation function is used in the 
output layer. As a loss function and an optimizer, the mean squared error and 
Adam \cite{Kingma2014} are adopted, respectively. Our deep CNN model 
has been implemented  using TensorFlow.\\

\begin{figure}[thbp]
    \centering
    \includegraphics[width=0.8\textwidth]{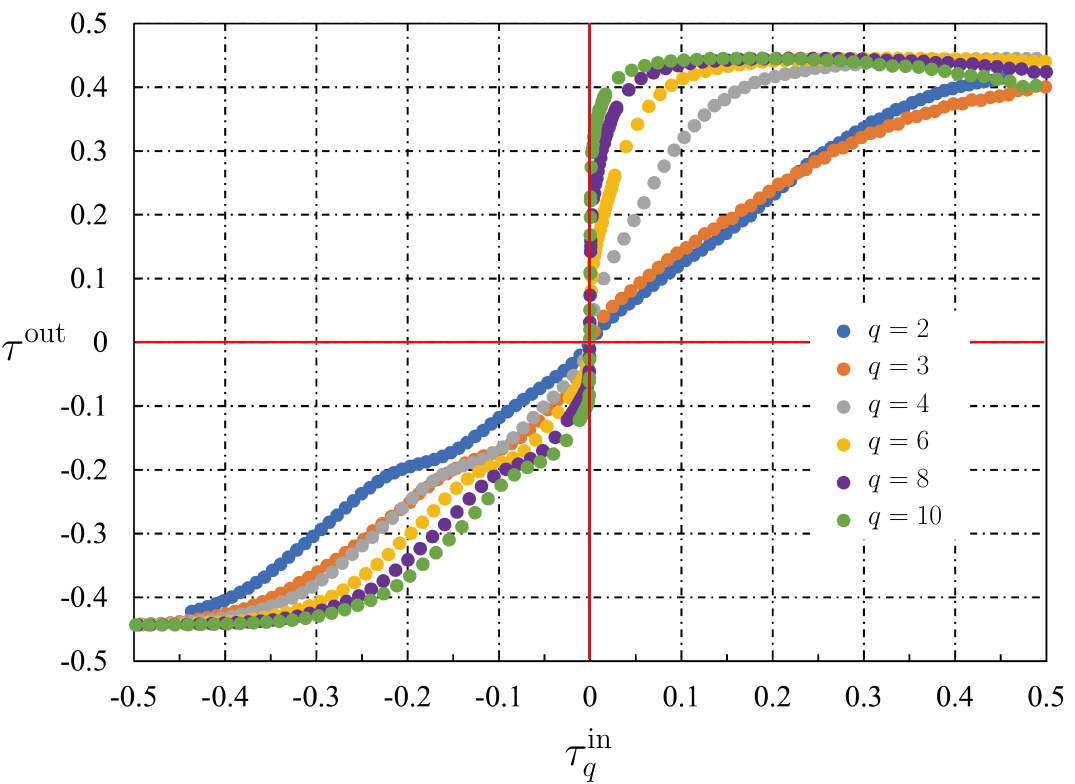}
    \caption{The relation of the normalized temperatures $\tau^{\rm in}$ \eqref{normalized} 
at which the input images of the $q$-state Potts models are generated and 
the corresponding normalized temperatures $\tau^{\rm out}$ \eqref{normalized} predicted by the CNN.
Each $\tau^{\rm out}$ is given by the mean value of the output data for the 6000 input images
generated at the same temperature. 
The transition points correspond to $\tau^{\rm in/out}=0$, which is well predicted by the CNN.
For more precise predicted values of the transition points, see Table~\ref{tc}. }
    \label{cnn_output}
\end{figure}

\noindent
4. {\it \large Results and Discussions}\,
Now we discuss the results. In Fig.~\ref{cnn_output}, we depict
the relationship of the temperatures $\tau^{\rm in}_q$ 
for the input images and the corresponding output temperatures 
$\tau^{\rm out}$ predicted by the CNN. Here, $\tau^{\rm in}_q$ 
and $\tau^{\rm out}$, respectively denote  
\begin{equation}
\tau^{\rm in}_q:=\frac{T-T_{{\rm c},q}}{T_{{\rm c},q}},\quad
\tau^{\rm out}:=\frac{T^{\rm CNN}-T_{\rm c}^{\rm CNN}}{T_{\rm c}^{\rm CNN}},
\label{normalized}
\end{equation}
where $T$ is the temperature at which input images are generated, 
$T_{{\rm c},q}$ is the transition temperature evaluated from the 
behavior of the internal energy  of the original $q$-state Potts model 
without binarization (due to finite-size effects, 
$T_{{\rm c},q}$ deviates from $T^{\infty}_{{\rm c},q}$ \eqref{transition} 
derived at the thermodynamic limit),  $T^{\rm CNN}$ denotes the average of 
the CNN outputs for the 6000 input samples generated at  $T$, 
and $T_{\rm c}^{\rm CNN}$ is the predicted critical
temperature corresponding to the Ising spin configurations 
at $T=T_{{\rm c},q=2}$. See Table~\ref{tc} for the detailed values.

\begin{table}[tbph]
    \centering
    \begin{tabular}{cccccccc}
    \hline
    \multicolumn{1}{c}{$q$} & \multicolumn{1}{c}{$T^{\ast}_{{\rm c},q}$} & 
    \multicolumn{1}{c}{$T_{{\rm c},q}$} & 
    \multicolumn{1}{c}{$T^{\infty}_{{\rm c},q}$ } & 
    \multicolumn{1}{c}{$E$} & 
    \multicolumn{1}{c}{$M$} & 
    \multicolumn{1}{c}{$d_{\rm f}$}&
     \multicolumn{1}{c}{transition type} \\
    \hline
    2 & $-$ & 1.1268& 1.1346  &   $-1.7216$ & $0.6726$, $-0.4800$  &11/8 &2nd\\
    3 & 0.9915 & 0.9906 & 0.9950  & $-1.7156$ &$0.6234$, $-0.6409$&17/12 &2nd\\
    4 & 0.9023 & 0.9066 & 0.9102  & $-1.7188$ &$0.7240$, $-0.5831$  &3/2 &2nd\\
    6 & 0.8042 & 0.8057 & 0.8076  & $-1.7107$ &$0.7036$, $-0.5839$  &$-$ &1st\\
    8 & 0.7429 & 0.7433 & 0.7449  & $-1.6945$ &$0.6546$, $-0.5686$  &$-$ &1st\\
    10& 0.6995  & 0.6994& 0.7012  & $-1.6587$ &$0.5813$, $-0.5211$  &$-$ &1st\\
    \hline
    \end{tabular}
    \caption{The transition temperatures of the $q$-state Potts models for various $q$.
 $T^{\ast}_{{\rm c},q}$, $E$ and $M$, respectively, denote the temperature,
internal energy, and magnetization of the  image
for which the CNN outputs $T_{\rm c}^{\rm CNN}=1.1209$. That is, these are
the quantities of the generated image that the CNN predicts 
to be at the transition point.  
$T_{{\rm c},q}$ is the transition temperature evaluated from the 
behavior of the internal energies for the $q$-state Potts model. $T^{\infty}_{{\rm c},q}$
is the transition temperature in the thermodynamic limit \eqref{transition}. Each datum
contains a numerical error in the last two digits.
The analytical values of the fractal dimension $d_{\rm f}$ ($q\le 4$)
in the thermodynamic limit are also listed.
}
\label{tc}
\end{table}
For comparison, the result for the Ising model ($q=2$) 
on which our CNN is trained is also depicted in the same figure. From this, one sees 
that our CNN is relatively well trained. The predicted temperatures 
for the input images generated at the same temperature, in general, depend on 
the model, except for the transition temperature $T=T_{{\rm c},q}$
 (i.e., $\tau_q^{\rm in}=0$).
This result is intuitively consistent, because the physical quantities 
such as the internal energy and magnetization \eqref{internal}, which are considered to be 
quantities that the CNN uses to make predictions \cite{KKT19}, generally depend on the model
(see below for a more quantitative discussion). On the other hand, for the images
generated at the transition temperatures $T=T_{{\rm c},q}$, the CNN outputs almost the same 
predicted temperatures as $T_{\rm c}^{\rm CNN}$. This result indicates that the Ising-trained
CNN can, surprisingly, detect the phase transition of the $q$-state Potts model,
despite the fact that the transition type,  physical quantities (e.g., $E$ and $M$), 
and geometric properties such as the fractal dimension of the spin-cluster boundaries, 
in general, are different from those of the Ising model, as shown in Table~\ref{tc}
(see also Fig. 2). 
The temperatures  $T^{\ast}_{{\rm c},q}$ 
of the input images from which the CNN outputs $T_{\rm c}^{\rm CNN}$ 
are listed in Table~\ref{tc}. (In other words, $T^{\ast}_{{\rm c},q}$
is the temperature of the generated image that the CNN predicts 
to be at the transition point.)
One finds that the CNN precisely detects the transition point
of the $q$-state Potts model, regardless of the type of transition.

\begin{figure}[tbp]
\centering
    \includegraphics[width=0.65\textwidth]{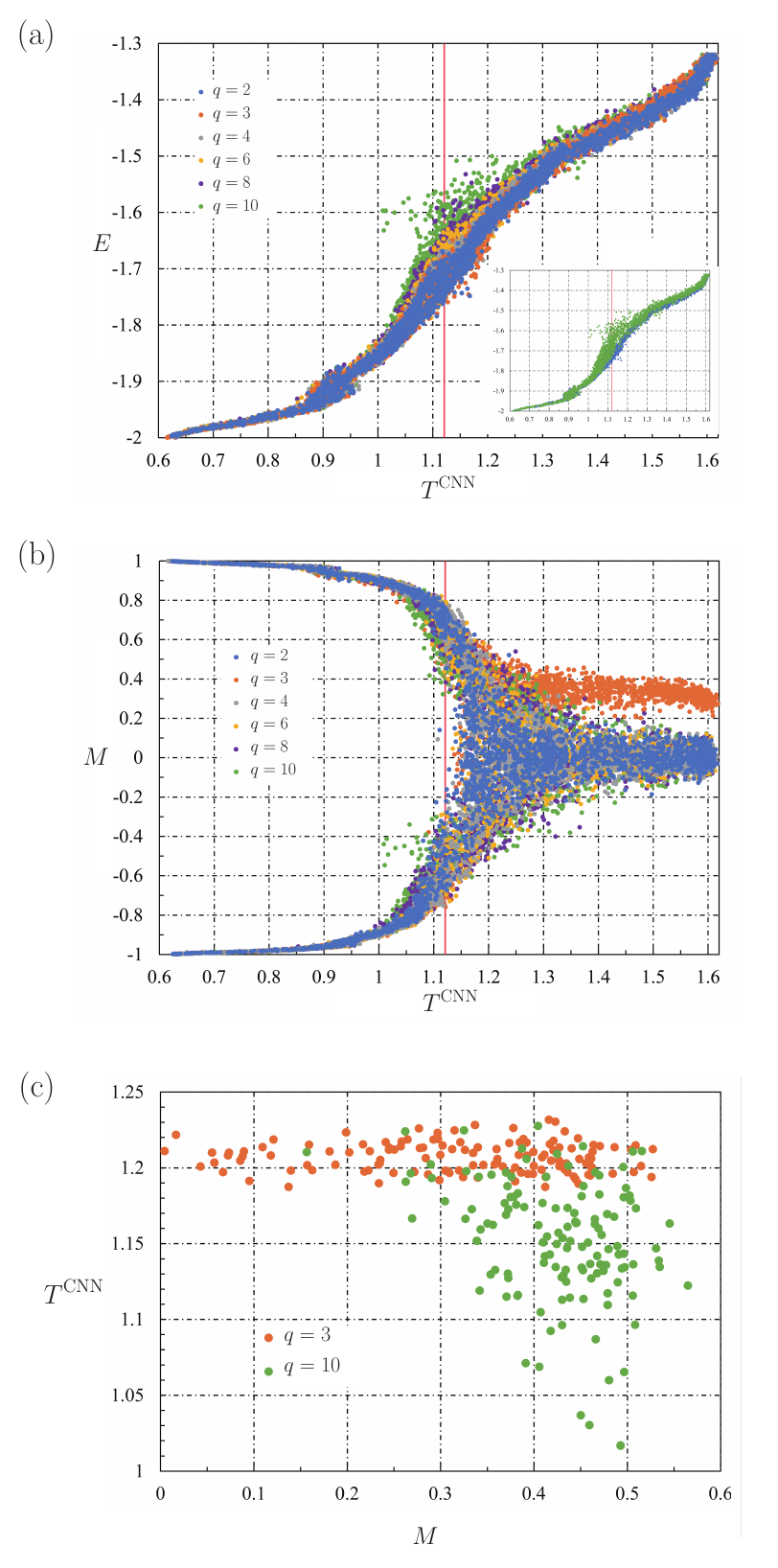}
    \caption{(a) The relation of the predicted temperatures $T^{\rm CNN}$ and the 
internal energies  $E$   for the binarized $q$-state Potts model.
(b) The relation of $T^{\rm CNN}$ and the magnetizations   $M$ for the
binarized model.
(c) The relation between $T^{\rm CNN}$ for the input images of the
3- and 10-state Potts models
with $E\in[-1.62,-1.6)$ and  $|M|$.}
    \label{energy-magnet}
\end{figure}

Next, let us consider how the CNN predicts the temperature for the input image.
In general, as pointed out in Ref. \cite{KKT19}, the CNN is 
considered to detect the phase transition 
by the magnetization or internal energy.
In Figs.~\ref{energy-magnet}(a) and (b), we depict the relation between the output
 temperatures of 
the CNN $T^{\rm CNN}$ and the internal energies $E$, and the magnetization 
$M$ (see Eq.~\eqref{internal} for their precise definitions).
One sees that above the transition point i.e., $T>T_{{\rm c},q}$,
the CNN outputs almost the same temperature if the $E$ of the
input images are the same, which indicates that the CNN predicts the temperatures 
based mainly  on  $E$ in the high-temperature region  $T>T_{{\rm c},q}$. Note that, 
in this region, the expectation value of $M$ 
does not depend on $T$. Correspondingly, the predicted temperatures
do not depend on $M$ either,  while, for the low-temperature region 
$T<T_{\rm c}$, 
the figure may indicate that the CNN outputs the temperature by both $E$ and $M$. In fact,
as explained below, a more detailed analysis using  principal component analysis indicates
that the CNN distinguishes between a system and that obtained by reversing all the spins.
Therefore, in the low-temperature region, CNN makes predictions 
based on $M$ (and possibly $E$ as well).
The dependence on $E$ or $M$ of the CNN output explains that the
predicted temperature $T^{\rm CNN}$ generally depends on the model, 
as shown in Fig.~\ref{cnn_output}.
However, in the vicinity of the transition point  $T=T^{\ast}_{{\rm c},q}$,  
the assumption that the CNN predicts the temperature based only on $E$ and $M$ 
does not account for the fact that our CNN can accurately predict the transition 
temperature. This is for the following two reasons. First, the internal energies $E$ (and
possibly $M$ because it appears to depend on the model) of the input images, 
which the CNN predicts to be at the transition points,
are different for each model  (see Table~\ref{tc}). 
Second, as depicted in Fig.~\ref{energy-magnet} (c) for $q=3$ and $q=10$, 
distributions of predicted temperatures for input images with approximately the same $E$ 
and $M$, are model dependent, which contradicts the assumption.
In conclusion, the Ising-trained CNN detects the transition point of the $q$-state
Potts model not only by the internal energies and magnetizations, but might detect it 
by more general properties such as complexities of spin clusters. 
It remains to be seen on which factors the CNN predicts the transition point.

\begin{figure}[tbp]
\centering
    \includegraphics[width=0.95\textwidth]{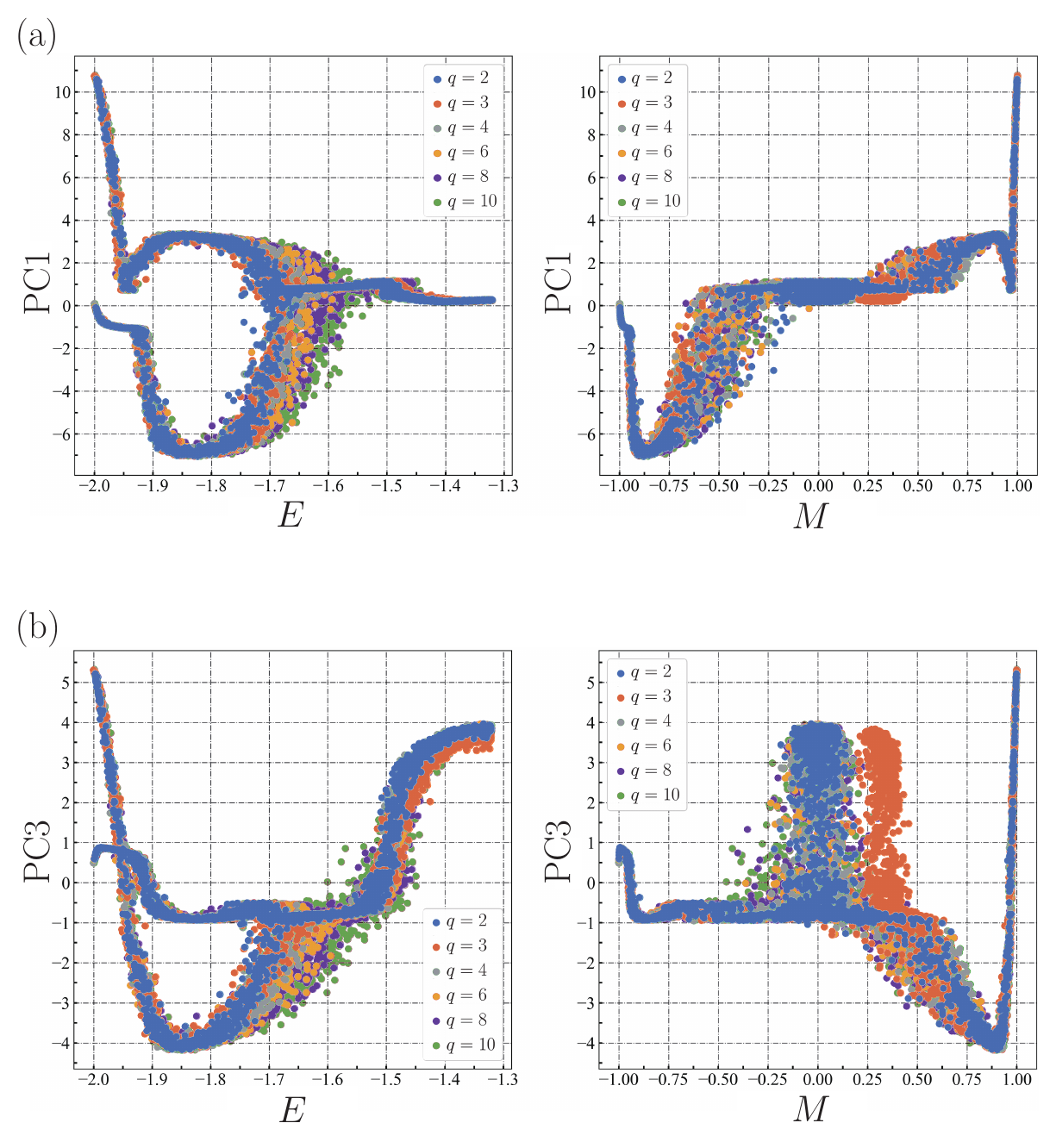}
    \caption{$E$ and $M$ dependences of the first principal component (a) and
the third principal component (b).
}
    \label{pc}
\end{figure}

To confirm the above results more concretely, we perform principal component analysis 
(PCA) on the last hidden layer consisting of 32 nodes in our CNN model (see Fig.\ref{cnn}). 
PCA is a 
multivariate analysis technique that uses an orthogonal transformation to rotate
 a multi-dimensional dataset (in our case 32 dimensions) so that each 
component of the transformed dataset is uncorrelated, and the 
components (called the principal components) are ordered in accordance 
with decreasing values of the variance  (i.e., the first principal
component has the greatest variance).
In Fig.~\ref{pc} (a), we depict the 
$E$ and  $M$ dependences of the
first principal component.
For the low-temperature region corresponding to $E<-1.75$ and 
$|M|> 0.75$ (see also Figs.~\ref{energy-magnet}(a) and (b)), the curves have finite gradients and
almost agree with those for the Ising model. Further, 
the  $M$ dependences are  single-valued functions, 
whereas the $E$ dependences are double-valued functions: the upper (lower) 
ones correspond to positive (negative) magnetizations. These behaviors
show that the CNN distinguishes  the difference between  images with positive 
magnetizations and those with negative magnetizations. Thus, in the low-temperature
region, the CNN outputs the temperature based mainly on $M$ and possibly $E$ as well.

One finds that the first principal component does not contribute to the high-temperature region
corresponding to $E>-1.5$ and $|M|<0.25$.
To detect how the CNN works in this region,
we also depict the behaviors of the third principal component in Fig.~\ref{pc} (b). 
In contrast to Fig.~\ref{pc} (a), in the high-temperature
region,
the $E$ dependences of the third principal component have finite gradients and
are consistent with that of the Ising model,
whereas the component does not depend on  $M$ in this region.
Thus, in the high-temperature region, the CNN predicts the temperature mainly based on $E$.

However, in the vicinity of the transition point corresponding to 
$-1.5<E<-1.7$ and $0.25<|M|<0.75$, the $E$ dependences of both the first and third components
explicitly depend on the spin state number $q$: on increasing  $q$, the curves gradually
shift to the right. 
Namely, the curves in this region depend on the model, and hence our CNN does not detect the 
transition  by $E$. On the other hand, the $M$ dependences in the vicinity of the
transition point have finite gradients and are similar to the behavior of the curve for the 
Ising model. The CNN might detect the transition point by $M$ or something 
related to that, but this is not particularly conclusive from these PCA analyses due to
the large variance.
As mentioned previously, further clarification of the properties used by
our CNN to detect transition points is a future issue.\\

\noindent
5. {\it \large Conclusion}\,
The Ising-trained deep CNN  can
precisely detect the phase transition of the $q$-state Potts model,
regardless of the type of transition. 
Our CNN model has not been trained on information about phases 
but is naively trained only by Ising 
spin configurations labeled with 
temperatures.
We find that, above the transition point, the deep CNN outputs the temperature 
mainly based on the internal energy, whereas, below the transition point, 
it outputs the temperature mainly based on the magnetization and possibly 
the internal energy. 
However, in the vicinity of the
transition point, the CNN predicts the temperature not only by the 
internal energy or the magnetization, but it may detect the transition 
points by more global features. In view of the fact that  NNs have 
been applied to detect more general types of phase transitions  such as
topological phase transitions \cite{OO16,ZK17,ZMK17,SRN17,BGM18,RKTM18,ZSZ18,SYZSZ18,VKK18,YAK18,
AMH19},  
a fundamental and quantitative investigation of how NNs capture global 
features at transition points is highly desired.

%
\section*{Acknowledgments}
The present work was partially supported by a Grant-in-Aid for Scientific
Research (C) No. 20K03793 from the Japan Society for the 
Promotion of Science.

\end{document}